\documentclass[journal=jpcafh,manuscript=article]{achemso}
\usepackage[utf8]{inputenc}

\usepackage{longtable}
\usepackage{multirow}
\usepackage{amsmath}
\usepackage{subfigure}
\usepackage[usenames,dvipsnames]{xcolor}
\usepackage{soul}
\usepackage{bm}

\usepackage{xr}
\usepackage{amssymb}
\usepackage{siunitx}
\usepackage{threeparttable}
\usepackage{multicol} \usepackage{wrapfig} \usepackage{enumitem}
\usepackage{bm} \usepackage{outlines} 
\usepackage{booktabs}
\usepackage{graphicx}
\usepackage[normalem]{ulem}
\usepackage{hyperref}
\usepackage{lineno}
\usepackage[version=4]{mhchem}
\usepackage[section]{placeins}

\makeatletter
\newcommand*{\addFileDependency}[1]{
  \typeout{(#1)}
  \@addtofilelist{#1}
  \IfFileExists{#1}{}{\typeout{No file #1.}}
}
\makeatother

\newcommand*{\myexternaldocument}[1]{%
    \externaldocument{#1}%
    \addFileDependency{#1.tex}%
    \addFileDependency{#1.aux}%
}
\myexternaldocument{si}

\author{JingChun Wang} \affiliation{Department of Chemistry,
  University of Basel, Klingelbergstrasse 80, CH-4056 Basel,
  Switzerland}

\author{Juan Carlos San Vicente Veliz} \affiliation{Department of
  Chemistry, University of Basel, Klingelbergstrasse 80, CH-4056
  Basel, Switzerland}

\author{Meenu Upadhyay} \affiliation{Department of Chemistry,
  University of Basel, Klingelbergstrasse 80, CH-4056 Basel,
  Switzerland}

\author{Markus Meuwly}\email{m.meuwly@unibas.ch}
\affiliation{Department of Chemistry, University of Basel,
  Klingelbergstrasse 80, CH-4056 Basel, Switzerland}

\title[]{High-Energy Reaction Dynamics of O$_{3}$}

\begin{document}
\date{\today}

\begin{abstract}
The high-temperature atom exchange and dissociation reaction dynamics
of the O($^3$P) + O$_2(^3\Sigma_g^{-} )$ system are investigated based
on a new reproducing kernel-based representation of high-level
multi-reference configuration interaction energies. Quasi-classical
trajectory (QCT) simulations find the experimentally measured negative
tempe-rature-dependence of the rate for the exchange reaction and
describe the experiments within error bars. Similarly, QCT simulations
for a recent potential energy surface (PES) at a comparable level of
quantum chemical theory reproduce the negative
$T-$dependence. Interestingly, both PESs feature a ``reef" structure
near dissociation which has been implicated to be responsible for a
positive $T-$dependence of the rate inconsistent with experiments. For
the dissociation reaction the $T-$dependence correctly captures that
known from experiments but underestimates the absolute rates by two
orders of magnitude. Accounting for an increased number of accessible
electronic states reduces this to one order of magnitude. A neural
network-based state-to-distribution model is constructed for both PESs
and shows good performance in predicting final translational,
vibrational, and rotational product state distributions. Such models
are valuable for future and more coarse-grained simulations of
reactive hypersonic gas flow.
\end{abstract}

\section{Introduction}
Molecular-level characterization of high-energy collisions in
shock-heated systems as they occur in atmospheric re-entry or
combustion is a challenging undertaking. Important processes under
such conditions include non-equilibrium excitations of internal
degrees of freedom following inelastic or reactive collisions and full
dissociation.\cite{park:1990} The gas flow in hypersonics is often in
a state of chemical and thermal non-equilibrium because the high
translational energy is rapidly converted to internal degrees of
freedom. The most prevalent chemical species in reactive air-flow
include N$_2$, O$_2$, NO and atomic nitrogen and oxygen. Hence, the
chemistry of oxygen-containing species is of particular
relevance.\cite{park:1990}\\

\noindent
For high-energy collisions a comprehensive characterization of the
state-to-state cross sections is an essential ingredient for
microscopic modeling hypersonic, reactive and rarefied gas flow at
high
temperatures.\cite{boyd:2015,boyd:2017,MM.hyper:2020,schouler:2020}
Using high-level electronic structure calculations for
full-dimensional, reactive and global potential energy surfaces (PESs)
together with their representation and dynamics simulations provides -
in principle - the necessary information for their
computation. However, characterizing {\it all} state-to-state reaction
cross sections and rates even for triatomics is a daunting task. This
is due to the large number of ro-vibrational reactant states of the
diatomic molecule $(\sim 10^4)$ that can combine any diatomic product
state (also $\sim 10^4$), which yields $\sim 10^8$ state-to-state
cross sections. For converging one such cross section from QCT
simulations, typically a minimum number of $10^4$ to $10^5$
simulations are required, which leads to $10^{12}$ to $10^{13}$ QCT
simulations that would have to be run at a given collision
energy. Depending on the range of collision energies ($\sim 5$ eV in
hypersonics) to be considered the total number of QCT simulations is
of the order of $10^{14}$ to $10^{15}$ which is neither possible nor
desirable.\\

\noindent
Over the past few years several machine learning-based treatments to
predict reaction outcomes have been
reported.\cite{MM.sts:2019,houston:2019,MM.nn:2020,MM.std:2021,komp:2022,coletti:2023,huang:2024}
One of the promising approaches is a state-to-distribution (STD)
model. For the N+O$_2$ $\rightarrow$ O+NO reaction an STD model
performed rather well (average $R^2 \sim 0.98$) to predict final state
distributions for arbitrary initial conditions when compared with
explicit QCT simulations.\cite{MM.std:2021} More importantly, thermal
rates $k(T)$ from using the STD model were also in excellent agreement
with rates determined from explicit QCT simulations over a wide range
of temperatures (1500 K to 20000 K).\cite{MM.std:2021}\\

\noindent
Dynamics studies of reactive processes require global and accurate
PESs. A particularly contested feature of the PES for the O($^3$P) +
O$_2(^3\Sigma_g^{-} )$ reaction concerns a so-called ``reef'' along
the minimum energy path for the atomic oxygen approaching the O$_2$
collision partner at long
range.\cite{fleurat:2003,babikov:2003,ayouz:2013,Dawes:2013,Li:2014,tyuterev:2014}
Although electronic structure methods at different levels of theory
find the reef\cite{pack:2002,schinke:2004,holka:2010}, removing it
yielded improved agreement between computed and measured thermal
rates.\cite{fleurat:2003} Also, the absence of a ``reef" has been
directly linked to a negative temperature dependence of the thermal
rate $k(T)$ from wavepacket calculations which is consistent with
observations.\cite{Li:2014} Several full-dimensional and reactive PESs
have been presented and used in dynamics studies in the
past.\cite{siebert:2001,siebert:2002,fleurat:2003,babikov:2003,ayouz:2013,ju:2007,mankodi:2017,Dawes:2011,Dawes:2013,varga:2017,Shu:2024}
In addition, more local PESs for spectroscopic studies have also been
presented.\cite{xie:2000,tyuterev:2013}\\

\noindent
For the O($^3$P) + O$_2(^3\Sigma_g^{-} )$ collision system multiple
reaction pathways are operative: 1) elastic and inelastic collisions,
2) the exchange of oxygen atoms (O$_{\rm A}$ + O$_{\rm B}$O$_{\rm C}$
$\rightarrow$ O$_{\rm B}$ + O$_{\rm A}$O$_{\rm C}$ or O$_{\rm C}$ +
O$_{\rm A}$O$_{\rm B}$) and 3) dissociation of the O$_2$ in the
entrance channel (O$_{\rm A}$ + O$_{\rm B}$O$_{\rm C}$ $\rightarrow$
O$_{\rm A}$ + O$_{\rm B}$ + O$_{\rm C}$). All these processes are
important in atmospheric chemistry and in reactive hypersonic
flow.\cite{boyd:2016,cacciatore:1978} In addition, the atom exchange
reaction (process 2) is known to exhibit particular isotope
effects.\cite{guo:2007,guo:2016} Both experimental and computational
studies were undertaken over the past decades to elucidate the
reaction dynamics and underlying intermolecular interactions governing
the
dynamics.\cite{siebert:2001,fleurat:2003,fleurat:2003.2,babikov:2003,fleurat:2004,Dawes:2011,ayouz:2013,Dawes:2013,tyuterev:2013,varga:2017,Shu:2024,lendvay:2019,tyuterev:2014}\\

\noindent
The present work reports the thermal rates for atom exchange and O$_2$
decomposition reactions using a reproducing kernel Hilbert space
(RKHS)-represented\cite{MM.rkhs:2017} PES based on multireference
configuration interaction with Davidson correction (MRCI+Q)
calculations for validation. In addition, such rates were also
determined for an earlier\cite{varga:2017} permutationally invariant
polynomial (PIP)-based PES for comparison and for better
characterizing the relationship between the overall shape of the PES
and the computed rates. Finally, for both PESs a neural network-based
(NN-based) STD model is generated which is useful for more
coarse-grained simulations of high-energy reactive airflow.\\

\noindent
This work is structured as follows. First the methods are
presented. Next, the PESs generated and used are characterized and the
results of QCT simulations for the atom exchange and atomization
reactions are described. Then, the STD models for the RKHS- and
PIP-based PESs are presented and discussed, followed by a discussion
of the results and conclusions.\\

\section{Methods}
\label{sec:2}

\subsection{The Reactive Potential Energy Surface}
The ground state PES for O$_3$ ($^1$A$'$) was constructed at the
MRCI+Q level\cite{Langhoff:1974,Werner:1988} with the augmented
Dunning-type correlation consistent polarized triple zeta
(aug-cc-pVTZ, AVTZ)\cite{Dunning:1989} basis set. This level of theory
has been found to adequately describe the electronic structure for
global and reactive PESs of triatomic C-, N-, and O-containing neutral
species.\cite{Lin:2016,MM.cno:2018,sanNO2} Furthermore, such a
treatment is consistent with previous work on thermal rates and final
state distributions for the [NNO], [OON], and [OOC] collision
systems\cite{MM.n3p:2022,MM.co2:2021,MM.no2:2020,MM.n2o:2020,MM.cno:2018}
which will allow consistent modeling and incorporation of the relevant
microscopic information - such as state-to-state or thermal rates - in
reaction networks encompassing all these species.\cite{boyd:2017} \\

\noindent
All electronic structure calculations were carried out in Jacobi
coordinates using the Molpro suite of codes and in $C_S$
symmetry.\cite{MOLPRO} The coordinates $R$, $r$, and $\theta$ are the
separation between one of the oxygen atoms and the center of mass of
O$_2$, the O$_2$ bond length, and the angle between the vectors
$\vec{R}$ and $\vec{r}$, respectively. Reference electronic structure
calculations were carried out on a grid comprising $R=[1.4, 1.6,
  1.8\cdots5.0, 5.25, 5.5, 6.0, 6.5, 7.0, 8.0, 9.5, 11.0, 13.0]$
a$_0$, $r=[1.1, 1.25, 1.5\cdots2.08, 2.19, 2.27, 2.35, 2.46\cdots3.21,
  3.79, 4.16]$ a$_0$, and $\theta=[169.796, 156.577, \\ 143.281,
  129.967, 116.647, 103.324, 90.100]^\circ$ with the other half of the
angles defined by symmetry. This leads to a total of 2269
geometrically valid and feasible ground-state geometries.\\

\noindent
For the MRCI calculations, multistate CASSCF(12,9) calculations were
carried out to correctly characterize the wave function which was then
used as the starting point for the MRCI+Q calculations. Preference of
MRCI+Q over conventional MRCI calculations was given because higher
level of accuracy can be achieved due to the mitigation of size
inconsistency of MRCI by introducing the Davidson
correction.\cite{rintelman:2005,werner:2011} It should be noted,
however, that including the Davidson correction in MRCI calculations
overestimates the dissociation energy for O$_3$ $\rightarrow$ O +
O$_2$ and underestimates the dissociation barrier
height\cite{fleurat:2003} at the 20 meV (0.5 kcal/mol) level. A total
of 8 states, two per spin state (singlet and doublet), and symmetry
group were included in the state-averaged calculations.\\

\noindent
As has been found previously for such calculations, either the CASSCF
or the MRCI+Q calculations may not always converge to the correct
electronic states or do not converge at all. Such energies were
removed and the grid was reconstructed and completed (``cleaned'')
using 1D RKHS $V(R;r)$ and/or 2D RKHS $V(R,r;\theta)$ interpolations
to evaluate the missing points. Using this cleaned grid the
3-dimensional RKHS representation was generated using the
kernel-toolkit.\cite{MM.rkhs:2017} The permutationally invariant,
reactive PES $V({\bf r}) = \sum_{i=1}^3 \omega_i(r_i)V_i({\bf r})$,
where $\mathbf{r}$ is the vector of three interatomic distances
($r_{\rm AB}$, $r_{\rm AC}$ and $r_{\rm BC}$), was constructed by
mixing the three possible channels O$_{\rm A}$O$_{\rm B}$+O$_{\rm C}$,
O$_{\rm A}$O$_{\rm C}$+O$_{\rm B}$, and O$_{\rm B}$O$_{\rm C}$+O$_{\rm
  A}$ using an exponential switching function
\begin{equation}\label{eq:mix}
    w_{i}(r_i)=\frac{e^{-(r_i / \rho_j})^{2}} {\sum_{j=1}^{3}
      e^{-(r_j / \rho_j})^{2}}
\end{equation}
for a given structure ${\bf r}$ of O$_3$. Using the ``mixing dataset''
(see below), the mixing parameters were determined by a grid-based
search to yield $\rho_{1}=\rho_{2}=\rho_{3}=1.40$ a$_0$. In total, the
RKHS-PES was constructed from 6300 reference energies. This compares
with 1686 energies determined at the XMS-CASPT2/maug-cc-pVTZ level of
theory from another, more recent global PES for
O$_3$.\cite{varga:2017}\\

\noindent
The ``mixing dataset'' was employed for optimizing the mixing
parameters $\rho_j$ in the switching function, see Eq~\ref{eq:mix}.
This grid was defined by $\theta=[30.0^{\circ},
  60.0^{\circ},90.0^{\circ},120.0^{\circ},150.0^{\circ}]$,
$r_{\rm AB}=[2.30, 2.40, 2.50, 2.55, 2.60, 2.65, 2.70, 2.75, 2.80, 2.85,
  2.90, 2.95, 3.00, 3.05, 3.10, 3.15, 3.20, 3.25,\\ 3.30, 3.40, 3.50,
  3.60, 3.70, 4.00]$ a$_0$ and $r_{\rm BC}=[2.30, 2.40, 2.50, 2.55, 2.60,
  2.65, 2.70, 2.75, 2.80,\\ 2.85, 2.90, 2.95, 3.00, 3.05, 3.10, 3.15,
  3.20, 3.25, 3.30, 3.40, 3.50, 3.60, 3.70, 4.00]$ a$_0$ to cover the
regions where the three channels overlap.\\

\noindent
Finally, an ``off-grid''
dataset was constructed to validate the overall performance of the
mixed RKHS-PES. The geometries were defined by $\theta=[20.0^{\circ},
  40.0^{\circ}, 80.0^{\circ}, 130.0^{\circ}, 160.0^{\circ}]$, and
$r_{\rm AB}$ and $r_{\rm BC}$ in the range of $[1.9, 5.1]$ a$_0$. None of the
off-grid points was used either in constructing the single-channel PES
or for optimizing the mixing parameters in the crossing regions. All
these reference calculations were again carried out at the MRCI+Q/AVTZ
level. For the ``mixing dataset'' energies larger than 300 kcal/mol
relative to full dissociation energy of O$_3$ were excluded.\\

\subsection{QCT Simulations}
For thermal rates (atom insertion and full dissociation), $5 \times
10^6$ independent QCT simulations were carried out for a given
temperature. Because $^{16}$O has nuclear spin $I=0$ only odd initial
$j-$values for O$_2$ are allowed.\cite{Li:2014} The maximum simulation
time was 75 ps or until the interatomic distance between the initial
diatomic O$_2$ was larger than 20 a$_0$ or the distance between the
O-atom and either of the atoms in the diatomic O$_2$ exceeded 24
a$_0$. Semiclassical initial
conditions were sampled from Boltzmann distributions of the total
angular momentum $J$, impact parameter $b$, collision energy $E_{\rm
  trans}$, and rovibrational states O$_2 (v,j)$. Thermal rates at
given temperature $T$ were then obtained from
\begin{equation}
 k(T) = g_e(T)\sqrt{\frac{8k_{\rm B}T}{\pi\mu}} \pi b^2_{\rm max}
 \frac{N_{r}}{N_{\rm tot}}.
\label{eq:thermal}
\end{equation}
Here, $g_e(T)$ is the electronic degeneracy factor, $\mu$ is the
reduced mass of the collision system, $k_{\rm B}$ is the Boltzmann
constant, and $N_r$ is the number of reactive trajectories (either
O-atom exchange or O$_3-$atomization). For the exchange
reaction,\cite{gross:1997,fleurat:2003,Dawes:2011,Dawes:2013}
$g_e(T)=3[5+3\exp(-277.6/T)+\exp(-325.9/T)]$ whereas for the full
dissociation reaction,\cite{fleurat:2003,Andrienko:2015,boyd:2016}
$g_e(T) = 1/27$ was adopted for all temperatures. The sampling
methodology was discussed in detail in
Ref.~\citenum{MM.cno:2018}. Statistical errors were quantified through
bootstrapping. For this, 10 batches of $10^5$ samples with 10 random
shuffles of the data (100 times of resamples in total) were used to
yield the expected thermal rates along with their standard
deviations.\\

\noindent
Construction of the STD model for the atom exchange reaction was based
on QCT simulations that were carried out on a grid of initial
conditions. These included $v=[0, 2,.., 12, 15,\\...,30, 34, 38, 42]$,
$j=[1, 5, 15, 29, 43, 57,..., 221, 235]$, and $E_{\rm col}=[0.5, 1.0,
  1.5,..., 5.0, 6.0, 7.0, 8.0]$ eV. State-specific QCT simulations
were carried out for a total of $10^5$ trajectories per initial
condition and stratified sampling was used for the impact parameter
$b=[0,b_{\rm max}=18.0\, {\rm a}_0]$.\\

\subsection{STD training and evaluation}
A neural network model was employed to predict collision outcomes
between an atom and a diatom.\cite{MM.std:2022} The model was trained
using 11 input features that characterized the reactant state, and it
produced 178 output nodes corresponding to the amplitudes of the three
product states. The NN architecture consisted of seven residual
blocks, each comprising two hidden layers. To enhance training
efficiency, the input features $x_{i}'$ were standardized
($\bar{x}'_{i}=0$, $\sigma'_{i}=1$) and the outputs were
normalized.\cite{lecun:2012} The training process minimized the
root-mean-squared deviation (RMSD) between the predicted and reference
quasi-classical trajectory (QCT) final state distributions, namely
$[P(E_{\rm trans}'), P(v'),P(j')]$. Since the NN outputs represent
probabilities and must remain non-negative even after normalization, a
softplus activation function was applied to the output layer.\\

\noindent
For training, the weights and biases of the NN were initialized using
the Glorot scheme,\cite{glorot2010understanding} and optimized with
the Adam algorithm,\cite{kingma2014adam} employing an exponentially
decaying learning rate. Training was performed using TensorFlow
1.0,\cite{abadi2016tensorflow} and the model parameters that yielded
the lowest loss on the validation set were selected for final
prediction. Overall, final state distributions from 2414 initial
conditions on a grid defined by $(0.05 \leq E'_{\rm c} < 5.0)$ eV with
$\Delta E'_{\rm c}=0.1$ eV, $(5.0 \leq E'_{\rm c} < 12.0)$ eV with
$\Delta E'_{\rm c}=0.2$ eV, and $(12.0 \leq E'_{\rm c} \leq 16.0)$ eV
with $\Delta E'_{\rm c}=0.4$ eV; $0 \leq v' \leq 42$ with step size 1;
and $0 \leq j' \leq 242$ (odd $j'-$values only) with step size $\Delta
j' = 6$ were generated.\\

\noindent
The results of the QCT simulations starting from 2414 initial
conditions were collected for generating the dataset. Simulations from
34 initial conditions yield insufficiently low reaction probabilities
$\sum_{v'=0}^{v_{\rm max}'} P(v') < 0.005$ for
convergence and were thus excluded from the training set. Initial
conditions for these low-probability final states are characterized by
low and extremely high $[E_{\rm trans}, v, j]$ values. This is because
for low initial values the atom exchange reaction is improbable to
occur whereas for the largest translational and/or internal energies,
atomization to 3 O($^3$P) is the dominant final state.\\

\noindent
The final dataset (final state distributions from 2380 initial
conditions) will be referred to as ``on-grid'' in the following. A
80:10:10 split of the dataset was randomly drawn for training,
validation and testing.\cite{MM.std:2022} The ``on-grid'' dataset is
to be distinguished from ``off-grid'' final state distributions that
originate from initial conditions which differ in any of the $[E_{\rm
    trans}, v, j]$ quantum numbers used for training. For additional
technical details, see Ref.\citenum{MM.std:2022}.\\

\section{Results}

\subsection{Construction and Validation of the PES}
\begin{figure}[h!]
  \centering
  \includegraphics[width=\textwidth]{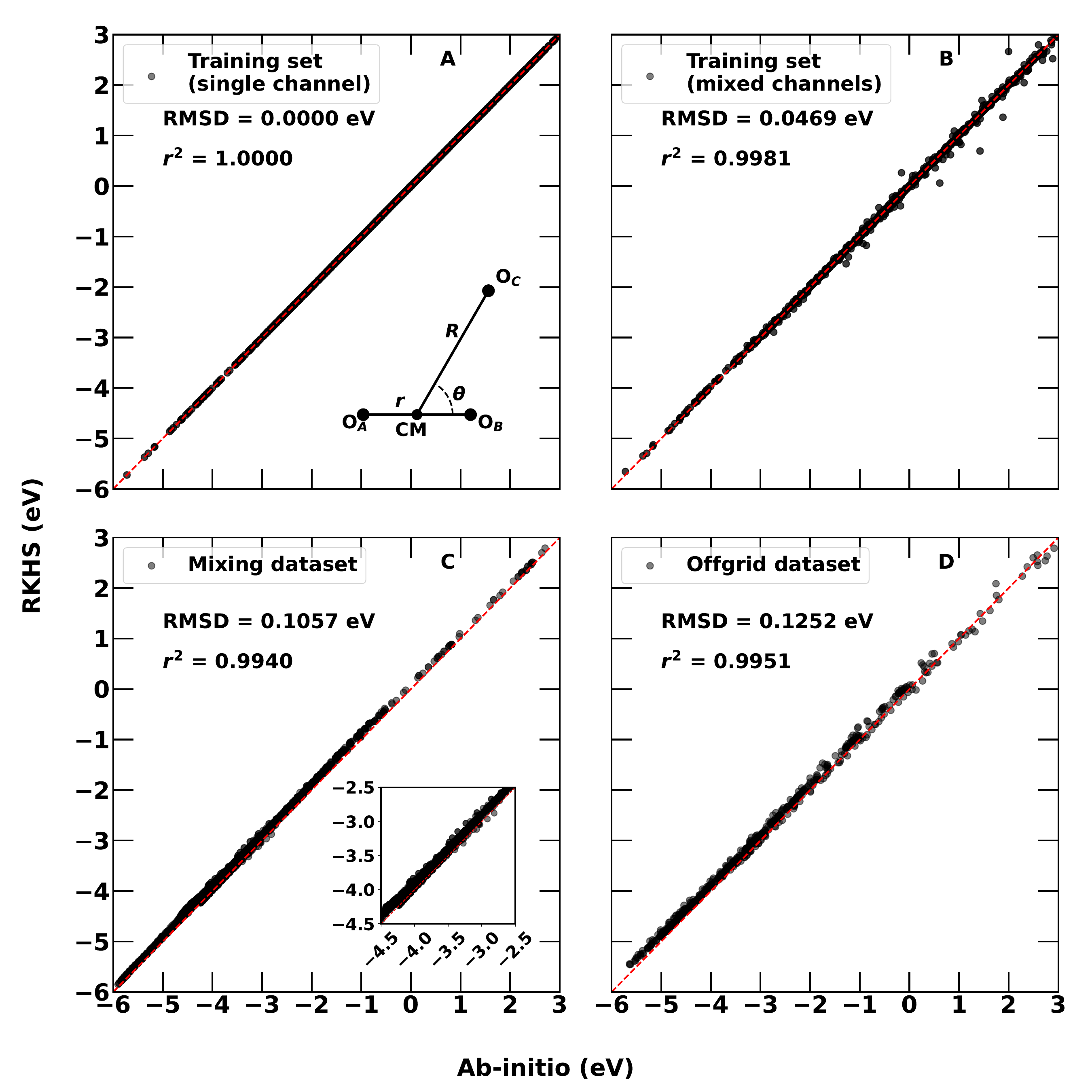}
    \caption{Validation of the RKHS-represented reactive,
      3-dimensional PES for O+O$_2$. Panel A: Correlation between
      single-channel RKHS representation and reference energies for
      on-grid training data. Panel B: Correlation between 3d mixed
      RKHS representation and reference energies for the on-grid
      training data. Panel C: Performance of the 3d mixed RKHS
      representation on the mixing dataset (see text). Panel D:
      Performance of the 3d mixed RKHS representation for 582 offgrid
      points. The RMSD between RKHS representation and reference data
      and the corresponding $r^2$ are given in each panel.}
    \label{fig:O3_PES_corr}
\end{figure}

First, the quality of the RKHS representation of the MRCI+Q/AVTZ
reference calculations is assessed. For the single-channel PES (Figure
\ref{fig:O3_PES_corr}A) the RKHS representation across 9 eV is of
exceptional quality (${\rm RMSD} < 10^{-5}$ eV; $r^2 = 1.0$). Mixing
the three PESs $V(r_{\rm AB}, R, \theta)$ and cyclic permutations
increases the RMSD to 0.047 eV with $r^2 = 0.9981$ for ``on-grid''
points, see Figure \ref{fig:O3_PES_corr}B. The low-energy part of the
energy range (between $-6$ and $-2$ eV) is still very accurately
represented whereas for energies 7 eV above the minimum energy a few
outliers appear. \\

\noindent
For the mixing dataset, Figure \ref{fig:O3_PES_corr}C, the RMSD is
0.11 eV with $r^2 = 0.9940$. No obvious outliers occur but the region
between --6 and --3 eV is somewhat widened, see inset. Finally,
validation of the reactive 3d PES on the offgrid data yields
$r^2=0.9951$ with an RMSD of 0.13 eV $(\sim 2.9 {\; \rm kcal/mol})$
which compares with an average RMSD of $\sim 4.5$ kcal/mol for a
recent PIP-PES across a comparable energy range (up to 200
kcal/mol).\cite{varga:2017}\\

\noindent
In constructing the global 3-dimensional reactive PES it was found
that certain geometries required particular attention. As a concrete
example, the point with $[R = 6.0 {\; \rm a}_0;\, r=3.79 {\; \rm
    a}_0;\, \theta = 169.796^\circ]$ is considered. Because for this
grid point the electronic structure calculations did not converge, the
energy for this point was inferred from converged energies at nearby
geometries. The extrapolated MRCI-predicted values is 0.4 kcal/mol
lower than the 1D RKHS predicted value, as in
Figure~\ref{sifig:comp-r3.79-refit}. The dashed line at
$169.796^{\circ}$ stands for the RKHS predictions without the extra
data of MRCI extrapolation at this point, while the solid grey line
for the RKHS predictions with the extra one point of training data.\\

\noindent
Next, the global reactive PES was characterized in terms of critical
points, shape and comparison with previous PESs which include
\begin{itemize}
\item PES1 based on multi-reference CI with singles and doubles
  excitation using the AVTZ basis set and represented as
  three-dimensional cubic splines,\cite{Schinke:2010}
\item PES2 (referred to as SSB), with energies obtained at the
  MRCI+Q/cc-pVQZ level of theory and represented as a cubic
  spline\cite{siebert:2001,fleurat:2003} based on $\sim 5000$
  reference energies,
\item PES3 (referred to as DLLJG) computed at the MRCI-F12/VQZ-F12
  level of theory, and using interpolating moving least squares
  (IMSL)\cite{Dawes:2011,Dawes:2013} to represent $\sim 2500$
  energies, and
\item PIP-PES constructed from $\sim 1700$ XMS-CASPT2 energies using
  the maug-cc-pVTZ basis set and represented as a permutationally
  invariant polynomial (PIP) surface.\cite{varga:2017}
\end{itemize}
In what follows, additional details are provided for each of these
PESs and the critical points from them are summarized in Table
\ref{tab:crit}. PES1 used electronic structure calculations for the
five lowest $^1$A$'$ states using the internally contracted
multireference configuration interaction method with single and double
excitations (MRD-CI) with the AVTZ basis set throughout.  The CI wave
functions were based on state-averaged (SA) CASSCF orbitals with 18
electrons in 12 orbitals (full-valence active space) and three fully
optimized closed-shell inner orbitals. The averaging includes the five
lowest $^1$A$'$ states with equal weights.\\

\noindent
PES2 was constructed at the level of internally contracted
MRCI+Q/cc-pVQZ based on CASSCF(12,9) reference wave
functions.\cite{siebert:2001,fleurat:2003} The subsequent MRCI
calculations included Davidson correction and PES2 was represented as
a three-dimensional splines which reproduce the input energies exactly
but no performance measures on off-grid points were provided.\\

\noindent
For PES3 the MRCI-F12/VQZ-F12 level of theory was
used.\cite{Dawes:2011,Dawes:2013} To improve convergence at large
separation, states correlating with the third molecular state of the
diatomic oxygen molecule were included in the dynamical weighting
SA-CASSCF(18,12) calculations, and 20 states (11 of symmetry $^1$A$'$
and 9 of symmetry $^1$A$'$$'$) were included. The MRCI-F12
calculations included 7 $^1$A$'$ reference states and Davidson
correction was applied. The data used for fitting PES3 included
energies within 2.6 eV (21000 cm$^{-1}$) of the global minimum and the
reported RMS fitting error was 2 cm$^{-1}$.\cite{Dawes:2013} A cap at
$\sim 60.0$ kcal/mol with respect to their reference energy was
applied for strongly repulsive arrangements.\\

\noindent
PES4\cite{varga:2017}, referred to as PES-PIP in the remainder of the
present work, used multi-state complete active space second-order
perturbation theory (XMS-CASPT2) with minimally augmented
correlation-consistent polarized valence triple-zeta(maug-cc-pVTZ)
basis set based on reference states from SA-CASSCF(12,9)
calculations. The 1s and 2s orbitals were fully optimized but were
kept doubly occupied in all configurations. To improve convergence to
the desired active space, restrictions were adopted on doubly occupied
orbitals. In all SA-CASSCF calculations, states were averaged with
dynamical weighting.  A level shift of 0.3 E$_h$ was applied to
mitigate intruder state errors, and the extended multi-state approach
was also used with fully invariant treatment of level shifts. The
overall accuracy of this PES: RMSE 2.9, 4.5, 8.4, 14.5, 6.1, and 26.2
kcal/mol for energies $<$100, 100-200, 200-500, 500-1000, 0-1000, and
$> 1000$ kcal/mol.\\

\noindent
Finally, there is also a very accurate representation of ic-MRCI/AV5Z
reference data primarily geared towards spectroscopic
applications. This analytical fit reproduces the ca. 5000 reference
energies to within a few cm$^{-1}$ over an energy range of 1.24 eV
(10000 cm$^{-1}$).\cite{tyuterev:2013}\\

\begin{table}[ht]
  \caption{Minima (MIN1, MIN2) for the RKHS and different PESs from
    the literature and from experiment. The transition state (TS1) was
    found using the Nudged Elastic Band
    (NEB)\cite{jonsson:2000,hammer:2016} method. MIN1 is the global
    minimum (open structure) of O$_3$, and MIN2 is the ring minimum
    $\sim 0.3$ eV above the O$_2$($^3 \Sigma_g^-$)+O($^3$P)
    dissociation threshold\cite{holka:2010}. The angle $\alpha$
    defined as $\angle$ $\rm{O_{B}O_{A}O_{C}}$($^\circ$) has atom
    O$_{\rm A}$ at its apex.}
    \begin{center}
    \begin{tabular}{|c|ccc|c|}
    \hline
    \hline
    $^{1}$A$'$  & $r_{e}^{\rm (O_{\rm A} O_{\rm B})}$(a$_0$) & $r_{e}^{\rm (O_{\rm A} O_{\rm C})}$(a$_0$) & $\alpha(\rm{O_{B}O_{A}O_{C}})$($^\circ$) & $\Delta E ({\rm kcal/mol})$ \\
    \hline
    {\bf MIN1} &                    &                     &                       &             \\
    RKHS-PES  & 2.43 & 2.43 & $116.1$ &  \\
    PES1\cite{Schinke:2010}  & 2.42 & 2.42 & $117.0$ &  \\
    PES2\cite{siebert:2001,fleurat:2003} & 2.41 & 2.41 & $116.8$ &  \\
    PES3\cite{Dawes:2011,Dawes:2013} & 2.40 & 2.40 & $116.8$ &  \\
    PIP-PES\cite{varga:2017} & 2.39 & 2.39 & $119.0$ &  \\
    Expt.\cite{tanaka:1970}  & 2.40 & 2.40 & $116.5$ &  \\
    \hline
    {\bf MIN2}                                                    &                    &                     &                       &             \\   
    RKHS-PES & 2.74 & 2.74 & $60.0$ & 29.2 \\
    PES2 & 2.72 & 2.72 & $60.0$ & 28.7 \\
    PES3 & 2.72 & 2.72 & $60.0$ & 30.8 \\
    PIP-PES & 2.69 & 2.69 & $60.0$ & 27.4 \\
    Theo.2\cite{gadzhiev:2013} & 2.73 & 2.73 & $60.0$ & 29.1 \\
    Theo.3\cite{kalemos:2008} & 2.72 & 2.72 & $60.0$ & 30.8 \\
    \hline
    {\bf TS}                                                       &                    &                     &                       &             \\
    RKHS-PES & 2.66 & 2.66 & $80.5$ & 48.0  \\
    PIP-PES                      & 2.63 & 2.63 & $81.0$ & 55.8 \\
    Theo.3\cite{kalemos:2008} & 2.76 & 2.76 & $87.0$ & 57.5 \\
    Theo.4\cite{chen:2011} & 2.66 & 2.66 & $84.0$ & 55.8 \\
    \hline
    \hline
  \end{tabular}
    \end{center}
  \label{tab:crit}
\end{table}

\noindent
Table \ref{tab:crit} summarizes and compares important characteristics
of the present PES compared with surfaces PES1 to PIP-PES from the
literature.\cite{Schinke:2010,siebert:2001,fleurat:2003,Dawes:2011,Dawes:2013,varga:2017}
The global minimum energy geometry (MIN1) on the RKHS-PES is an
isosceles triangle with $r_{\rm AB} = r_{\rm BC} = 2.43$ a$_0$ and an
apex angle of $116.1^\circ$ at atom O$_{\rm B}$. There is a second,
local minimum (MIN2) with an equilateral triangle structure and
separations 2.74 a$_0$ between all oxygen atoms. The energy difference
between these two structures on the RKHS PES is 29.2 kcal/mol which
compares with 30.7 kcal/mol from minimization of both structures at
the MRCI+Q/AVTZ level of theory, and 29.0 kcal/mol on the
single-channel RKHS-PES. The difference between the optimizations on
the RKHS-PES and the electronic structure calculations are due to
slightly different geometries. Optimizations using quantum chemistry
compress the bond separations from 2.43 a$_0$ to 2.42 a$_0$ and
increase the angle from $116.1^\circ$ to $116.8^\circ$. Using the
larger AVQZ basis set this energy difference increases by 0.5
kcal/mol.\\

\noindent
For MIN1 the optimized structure from the RKHS-PES is consistent with
previous work although differences of $\sim 0.02$ a$_0$ in bond
lengths and up to $3^\circ$ can be found, see top block in Table
\ref{tab:crit}. The experimentally reported equilibrium structure has
$r_e = 2.40$ a$_0$ and $\theta_e = 116.5^\circ$.\cite{tanaka:1970} The
structure of the higher-lying minimum, MIN2, was also determined in
previous work and yields bond lengths within $\sim 0.05$ a$_0$ of the
present calculations. MIN2 is 29.2 kcal/mol from the RKHS-PES which
compares with a range of 27.4 kcal/mol to 30.8 kcal/mol on earlier
PESs, calculated at comparable levels of quantum chemical theory but
not represented as a RKHS. Finally, the TS between MIN1 and MIN2
features bond lengths of $2.66$ a$_0$ and a bond angle of
$80.5^{\circ}$ with an energy of 48.0 kcal/mol above MIN1. This
compares with results from the literature that report bond lengths
around $\sim 2.7$ a$_0$ with an equilibrium angle between $81^\circ$
and $87^\circ$ and energies $\sim 60$ kcal/mol above MIN1 but with
considerable variations for the PESs that reported this TS. \\

\begin{figure}[h!]
    \centering
    \includegraphics[width=\linewidth]{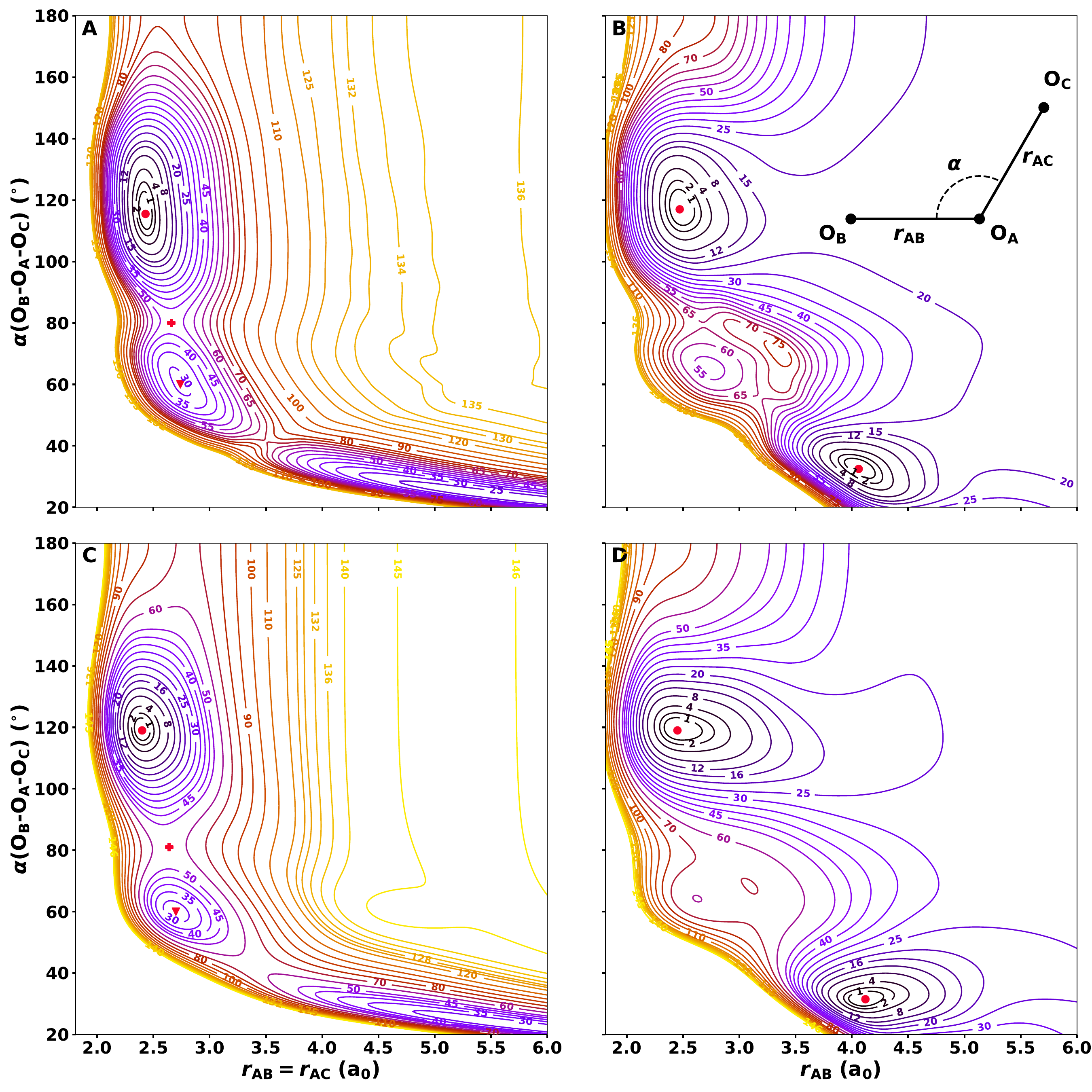}
    \caption{Contours for the O$_3$ RKHS-PES and PIP-PES\cite{varga:2017}. 
    Energies are in kcal/mol with the global minimum (MIN1) energy as the
      zero of energy. Panels A/B and C/D for the RKHS-PES and PIP-PES,
      respectively. In panels A and C the two bond lengths forming the
      angle $\alpha$(O$_{\rm B}$-O$_{\rm A}$-O$_{\rm C}$) are
      identical. In panels B and D the bond length $r_{\rm BC} =
      2.283$ a$_0$ while scanning the other bond length. Red circles,
      triangles and crosses designate MIN1 (global minimum), MIN2
      (tight minimum), and TS1.}
    \label{fig:Contour-Constr}
\end{figure}

\noindent
Another property of the PES that impacts the high-energy reaction
dynamics of O$_3 (^1 {\rm A}_1)$ is the dissociation energy $D_e$ to
form O$_2 (^3 \Sigma_{\rm g}^{-})$ + O$(^3 {\rm P})$. There is no
direct measurement of $D_e$ but analysis of thermochemical data yields
an estimate of $D_e^{\rm expt.} = 26.1 \pm 0.4$ kcal/mol that is
typically used for
comparison.\cite{janaf,muller:1998,xie:2000,siebert:2001,tyuterev:2013,varga:2017}
The RKHS-PES yields 21.7 kcal/mol which compares with 19.8 kcal/mol
from minimized structures for O$_3$ and O$_2$ at the MRCI+Q/AVTZ level
of theory. Using the AVQZ basis set this increases to $D_{\rm e} =
21.4$ kcal/mol. This is consistent with early work that reported a
pronounced dependence of $D_e$ on the basis set size. Compared with
the complete basis set (CBS) limit at the CCSD(T) level, using an AVTZ
basis set underestimates $D_e^{\rm expt.}$ by $\sim 3$ kcal/mol and
the CBS limit is lower by 1.3 kcal/mol compared with the predicted
$D_e$.\cite{muller:1998} Hence, CCSD(T)/AVTZ calculations
underestimate the assumed $D_e^{\rm expt.}$ by 4.2
kcal/mol.\cite{muller:1998} Subsequent icMRCI+Q calculations using the
cc-pVQZ basis set reported\cite{xie:2000} a value of 24.2 kcal/mol,
which is larger than the value of 22.6 kcal/mol from CCSD(T)/cc-pVQZ
calculations.\cite{muller:1998} Selected more recent calculations
reported values of 23.7 kcal/mol
(icMRCI+Q/cc-pVQZ)\cite{siebert:2001}, 23.0 to 25.8 kcal/mol (icMRCI+Q
with AVTZ to AV6Z basis set)\cite{tyuterev:2013}, and 26.7 kcal/mol
($D_e^{\rm expt.}$ included in fit; the XMA-CASPT2 value is 36.0
kcal/mol)\cite{varga:2017}. Hence, all these calculations confirm that
using the AVTZ basis set underestimates $D_e^{\rm expt.}$ by up to 4
kcal/mol.\\

\noindent
The shapes of the present RKHS-PES with those from the literature are
compared in Figures \ref{fig:Contour-Constr} and
\ref{sifig:cont-compare}. The RKHS-PES and PES-PIP shown in Figures
\ref{fig:Contour-Constr}A/C and B/D are deceptively similar from a
visual comparison. The only major difference concerns the region
around $\theta = 120^\circ$ and $r_{\rm AB} > 3.0$ a$_0$ for which the
RKHS-PES is tighter and does not extend to longer separations compared
with PES-PIP. Given the rather different underlying grid and
representation strategy (RKHS vs. PIP) such close agreement is
notable. It is anticipated that simulations using these two PESs yield
comparable results as will be discussed in the next subsection. Figure
\ref{sifig:cont-compare} compares $V(r_{\rm AB},r_{\rm BC})$ for the
RKHS-PES, PES-PIP and PES3. Again, the three PESs share common
topographies in particular in the bound state region. For high-energy
regions (yellow isocontours) the RKHS-PES and PES-PIP behave in a
comparable fashion whereas PES3 features rather sharp edges.\\

\begin{figure}[h!]
    \centering
    \includegraphics[width=\textwidth]{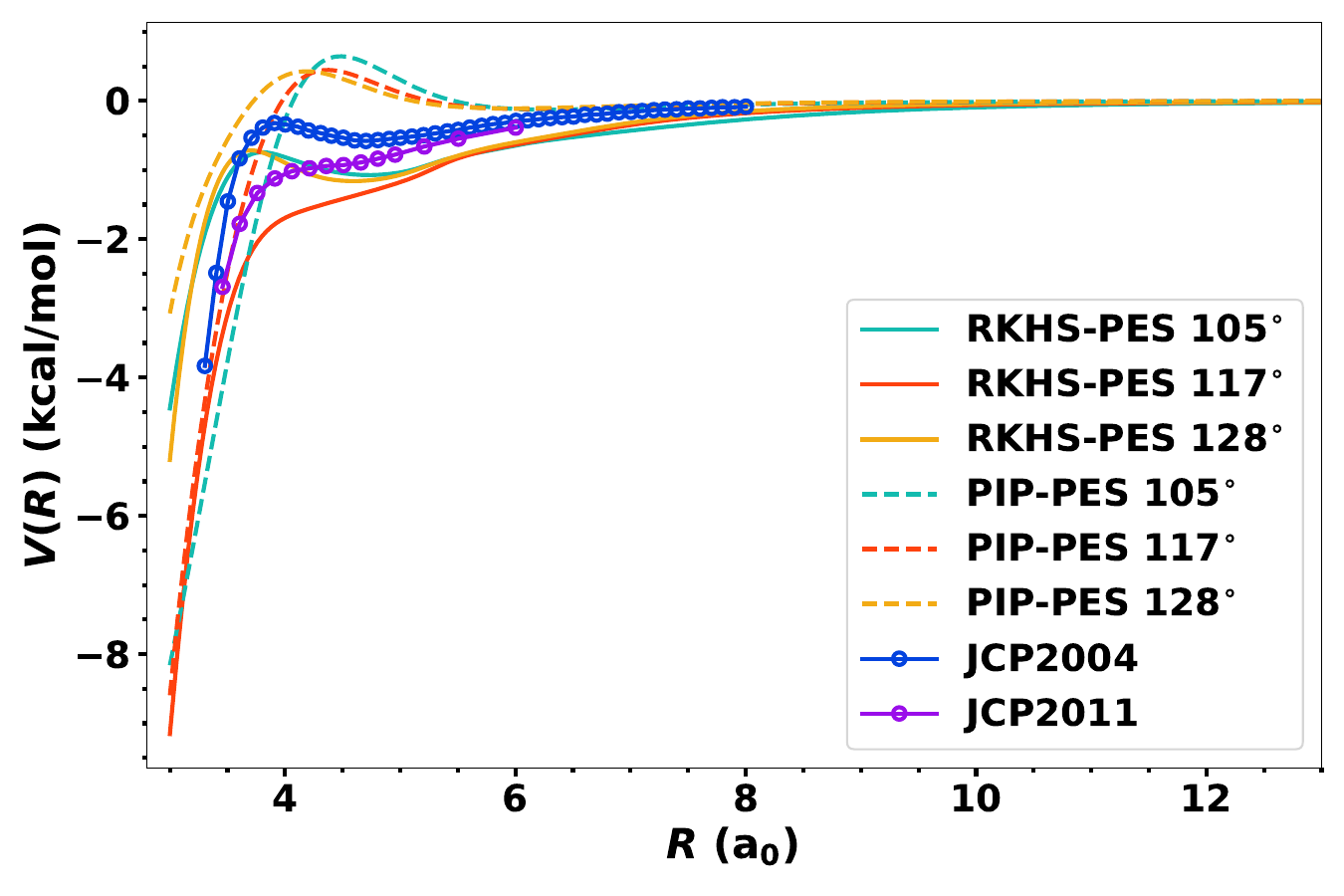}
    \caption{Comparison of the $R-$dependence $V(R)$ of the RKHS-PES
      (solid lines), the PIP-PES\cite{varga:2017} (dashed lines),
      PES2\cite{schinke:2004} (blue circles), and
      PES3\cite{Dawes:2011} (magenta circles). PES2 features the
      ``reef'' whereas PES3 was designed to have the ``reef''
      removed. The Jacobi angles $\theta$ for the RKHS and PIP-PES are
      $105^\circ$ (green), $117^\circ$ (red), and $128^\circ$
      (orange), respectively, see Figure
      \ref{fig:O3_PES_corr}A. Depending on the angle considered, the
      RKHS-PES features a ``reef'' or not whereas the PIP-PES has a
      ``reef'' for all three values of $\theta$.}
    \label{fig:pes-comp-m}
\end{figure}

\noindent
Finally, it is of interest to compare 1-dimensional cuts in the
near-dissociation region with a particular emphasis on the ``reef''
structure that has spurred intense
discussions.\cite{fleurat:2003,babikov:2003,ayouz:2013,Dawes:2013,Li:2014,tyuterev:2014,pack:2002,schinke:2004,holka:2010}
Figure \ref{fig:pes-comp-m} compares 1-dimensional scans along the
O+O$_2$ dissociation coordinate $R$ within $< 10$ kcal/mol of the
dissociation energy. Both, the RKHS-PES (solid lines) and PES-PIP
(dashed lines) feature more or less pronounced (submerged) barriers
depending on the angle of approach $( 105^\circ < \theta < 128^\circ
)$. On the other hand, the green solid line do not display this
feature. The particular relevance of the ``reef'' is the fact that a
positive temperature dependence for $k(T)$ has been implicated to
arise from PESs featuring such a reef whereas reef-free PES were found
to lead to a negative $T-$dependence which is consistent with
experiments.\cite{fleurat:2003,babikov:2003,ayouz:2013,Dawes:2013,Li:2014,tyuterev:2014,pack:2002,schinke:2004,holka:2010}
As will be shown below, $k^{\rm exch}(T)$ from both, RKHS-PES and
PES-PIP, lead to a negative $T-$dependence in agreement with
experiment despite the reefs that are present in Figure
\ref{fig:pes-comp-m}.\\

\subsection{The Exchange and Full Dissociation Reactions}
\begin{figure}
    \centering
    \includegraphics[width=\textwidth]{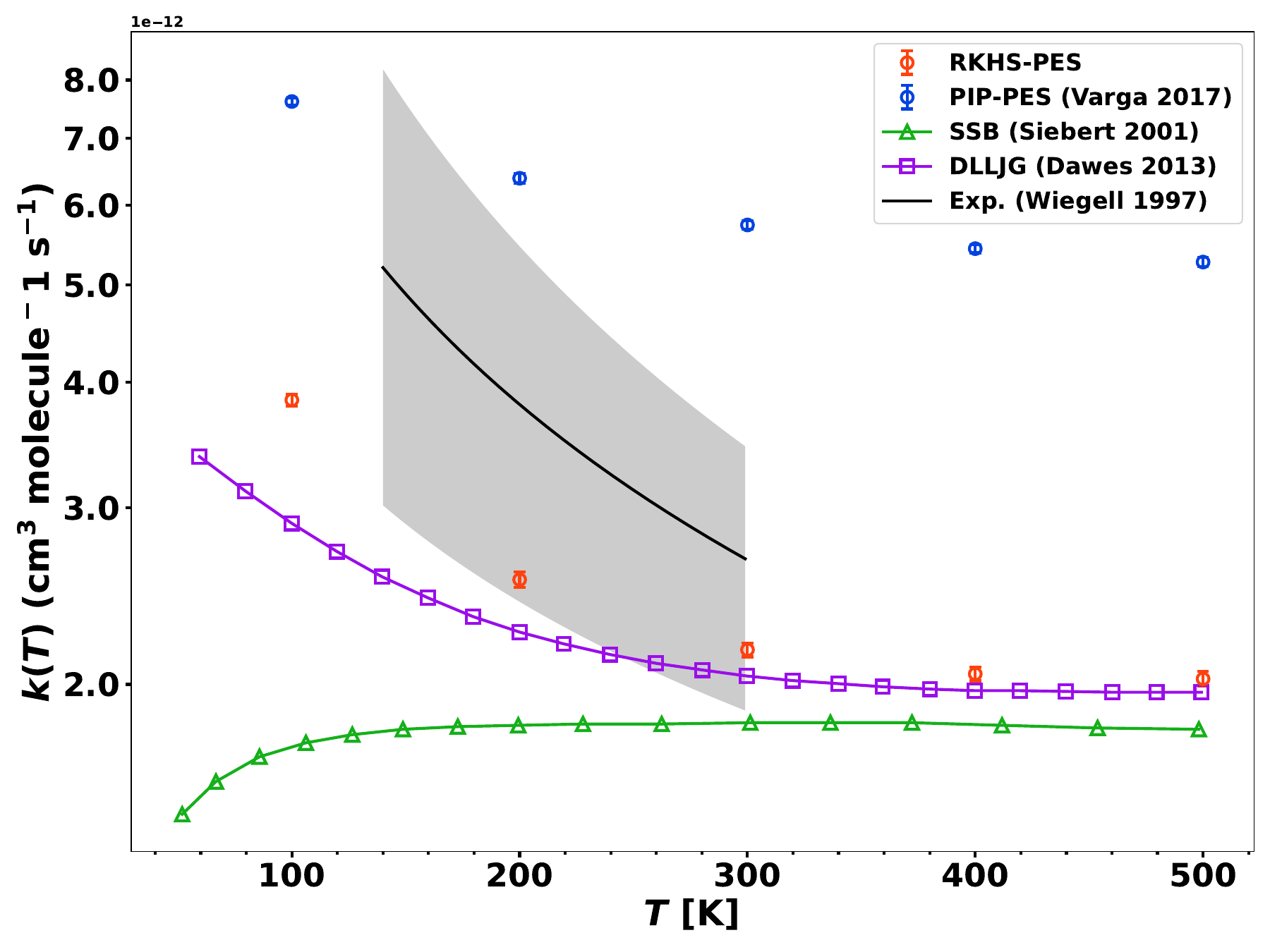}
    \caption{Comparison of $k^{\rm exch}(T)$ for the O$_2$ + O
      $\rightarrow$ O$_2$ + O exchange reaction from the present QCT
      simulations using the RKHS (orange circles) and
      PIP\cite{varga:2017} (blue circles) PESs and earlier Jacobi
      coordinate-based quantum wave packet method simulations using
      the DLLJG\cite{Dawes:2013,Li:2014} (violet) and
      SSB\cite{siebert:2001,babikov:2003} PESs (green). The
      experimentally measured rates (black line) are shown together
      with the reported uncertainties (grey
      background).\cite{fleurat:2003.2,wiegell:1997}}
    \label{fig:O3-exch-all}
\end{figure}

\begin{figure}[h]
    \centering
    \includegraphics[width=\textwidth]{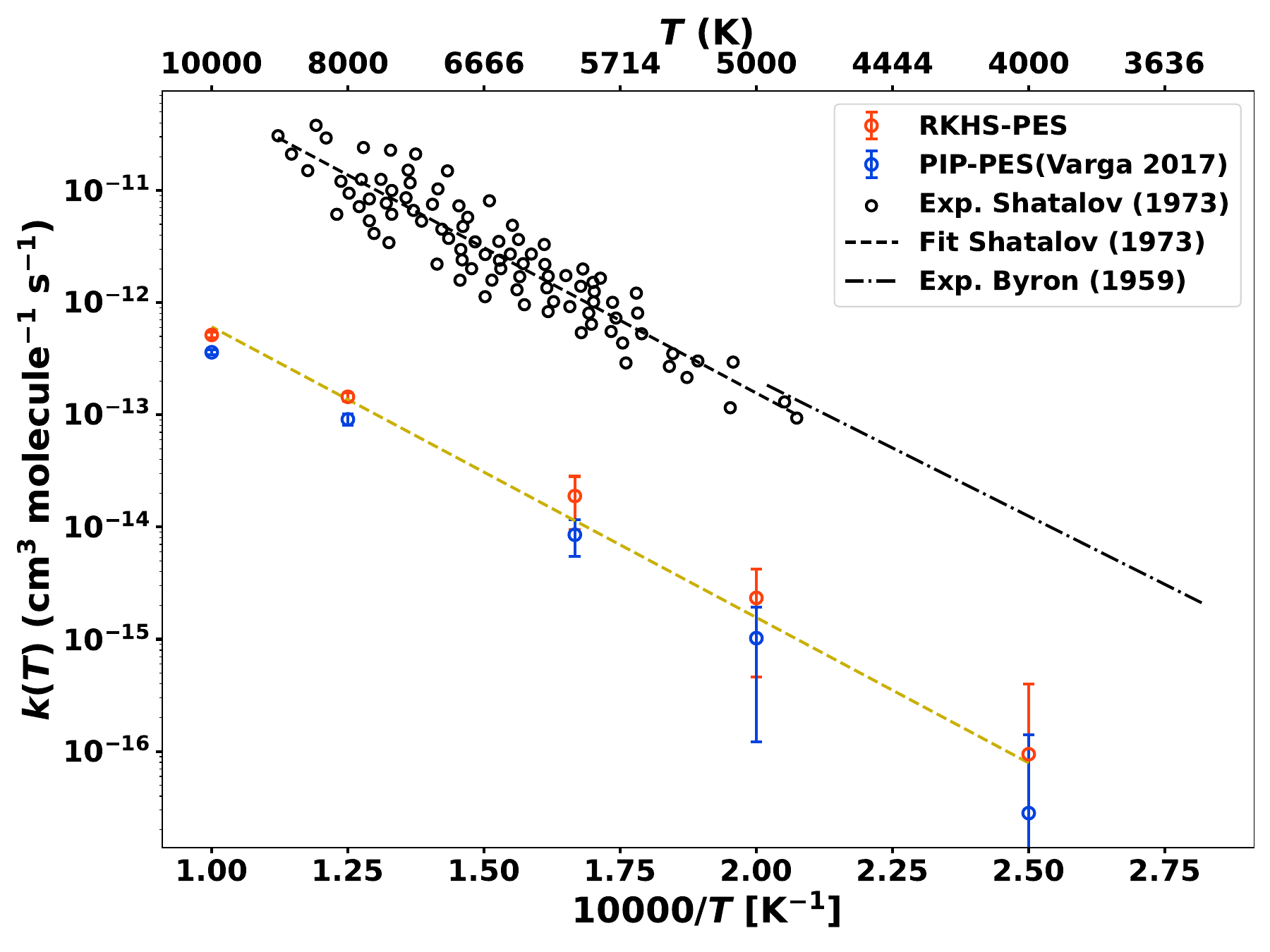}
    \caption{Rates $k^{\rm diss}(T)$ for the dissociation reaction
      from QCT simulations. Simulation results using the RKHS- and
      PIP-PESs are the red and blue symbols including error bars from
      bootstrapping. The measurements are in
      black\cite{Byron:1959,Shatalov:1973} together with a linear
      regression (dashed black line). To visually underscore the
      correct $T-$dependence of the simulations, the dashed black line
      was shifted (dashed red line) to best match the computed data.}
    \label{fig:O3-dis}
\end{figure}

Next, thermal rates $k^{\rm exch}(T)$ and $k^{\rm diss}(T)$ for the
atom exchange (O$_{\rm A}$ + O$_{\rm B}$O$_{\rm C}$ $\rightarrow$
O$_{\rm B}$ + O$_{\rm A}$O$_{\rm C}$ or O$_{\rm C}$ + O$_{\rm
  A}$O$_{\rm B}$) and dissociation O($^3$P) + O$_2( ^3\Sigma_g^{-} )$
$\rightarrow$ 3O($^3$P) reactions, respectively, are discussed, see
Figures \ref{fig:O3-exch-all} and \ref{fig:O3-dis}.\\

\noindent
{\it Exchange Reaction:} Thermal rates for the atom exchange reaction
(O$_{\rm A}$ + O$_{\rm B}$O$_{\rm C}$ $\rightarrow$ O$_{\rm B}$ +
O$_{\rm A}$O$_{\rm C}$ or O$_{\rm C}$ + O$_{\rm A}$O$_{\rm B}$) from
simulations at 100, 200, 300, 400 and 500 K together with associated
error bars from bootstrapping using the two reactive PESs, the RKHS
PES and PIP-PES, are given in Figure \ref{fig:O3-exch-all}. The
thermal rates $k^{\rm exch}(T)$ including the electronic degeneracy
factor\cite{gross:1997,fleurat:2003,Dawes:2011,Dawes:2013}
$g_e(T)=3[5+3\exp(-277.6/T)+\exp(-325.9/T)]$ from using the RKHS and
PIP-PESs are reported as red and blue symbols in Figure
\ref{fig:O3-exch-all}, respectively. Both PESs lead to a negative
$T-$dependence for $k(T)$, which is consistent with the experiments
(black line with shaded area indicating measurement
errors).\cite{wiegell:1997} In addition, earlier results from quantum
wavepacket simulations using the SSB (green)\cite{siebert:2001} and
DLLJG (lila)\cite{Dawes:2013} PESs are shown, respectively.\\

\noindent
The results indicate that the QCT simulations using the RKHS-PES yield
$k^{\rm exch}(T)$ consistent with experiments within the error bars
but underestimate the reported values from experiment somewhat. This
is also the case for the wavepacket simulations using the DLLJG PES
(PES3). QCT simulations using the PIP-PES yield the correct
$T-$dependence but overestimate $k^{\rm exch,exp}(T)$
somewhat. Finally, using the SSB PES (PES2) leads to a positive
$T-$dependence which is not what the measurements report. It is of
interest to mention that PES3 had the ``reef'' removed (see Figure
\ref{fig:pes-comp-m}) whereas the RKHS and PIP-PESs both feature
``reefs'' as does PES2. Nevertheless, the $T-$dependence of the RKHS
and PIP-PESs follows that from the measurements which indicates that
the presence or absence of the ``reef'' is not directly related to
capturing the correct $R-$dependence known from the experiments.\\

\noindent
It is of interest to note that the fraction of reactive trajectories
ranges from $\sim 6 \%$ to $\sim 10 \%$ in the temperature range
considered. The majority of the trajectories feature inelastic
scattering with changes of ($v_f, j_f$) and $E_{\rm trans}$. At $T =
100$ K, 25\% of the trajectories are elastic, and 65\% of the
trajectories are inelastic. The number of trajectories that
remain in the O$_3$ state for longer than 75 ps is 2766, 866, 486,
339, 252 out of $5\times 10^6$ for $T=$ 100, 200, 300, 400, 500 K,
respectively.\\

\noindent
{\it Dissociation Reaction:} For the dissociation reaction O$_2 (^3
\Sigma_{\rm g}^{-})$ + O$(^3 {\rm P})$ $\rightarrow$ 3 O$(^3 {\rm P})$
the thermal rates $k^{\rm diss}(T)$ are reported in
Figure~\ref{fig:O3-dis}. The calculated dissociation rates between
1000 K and 20000 K for both the RKHS-PES and PIP-PES from QCT
simulations follow the negative $T-$dependence reported from
measurements for $4000 < T < 10000$ K.\cite{Byron:1959,Shatalov:1973}
Fitting the measured dissociation rates to a linear regression (dashed
black line) and shifting to best overlap with the computed rates
(dashed red line) demonstrates that the simulations recover the
correct $T-$dependence (slope) of $k^{\rm diss}(T)$. It should be
noted that for both PESs the number of reactive trajectories decreases
significantly as $T$ decreases, see
Tables~\ref{sitab:ratio_disso_RKHS} and
Table~\ref{sitab:ratio_disso_PIP}. The number of trajectories that
remain as O$_3$ for longer than 75 ps are 78, 23, 10, 2, 2, 3, 1, 2,
0, 1 out of $5\times 10^6$ for $T= 1000$, 2000, 3000, 4000, 5000,
6000, 8000, 10000, 15000, and 20000 K, respectively.\\

\noindent
Although the $T-$dependence for $k^{\rm diss}(T)$ is correctly
described compared with the measurements, the dissociation rates are
too small by $\sim 2$ orders of magnitude. There are several factors
that potentially influence this finding. First, the dissociation
reaction probes, inter alia, the strength $D_e$ of the O$_2$
bond. Geometry optimization of O$_2$ at the MRCI+Q level of theory
using the AVTZ and AVQZ basis sets yield 113.5 and 116.4 kcal/mol,
respectively, which compare with 115.0 kcal/mol from the RKHS-PES a
value of $D_{\rm e} ^{\rm expt.} = 120.6$ kcal/mol from combining
results from measurements.\cite{bytautas:2010,ruscic:2004,creek:1975}
On the other hand, when conceiving the PIP-PES the O$_2$ dissociation
energy was included in the fit and the value is $D_e = 120.6$
kcal/mol. Given that the dissociation rates from QCT simulations using
the RKHS and PIP-PESs are within a factor of $\sim 3$but the value of
$D_e^{\rm O_2}$ differs by $\sim 5$ kcal/mol suggests that $k^{\rm
  diss}(T)$ is not particularly sensitive to the value of $D_e^{\rm
  O_2}$.\\
  
\noindent
Secondly, the electronic degeneracy factor was $g_{\rm e} = 1/27$
throughout, as was suggested in previous
work\cite{fleurat:2003,Andrienko:2015,boyd:2015} on high-temperature
simulations for O$_3$. Increasing the degeneracy factor - which
amounts to including higher lying electronic states that become
populated at the collision energies considered in the present work -
will increase the dissociation rates and improve the quantitative
agreement with experiments. Reconsidering the electronic degeneracy
has already been proposed, in particular for high collision
energies.\cite{boyd:2016} Based on earlier work, in the
high-temperature limit the electronic degeneracy should increase to at
least $g_{\rm e} = 16/27$ which still includes population of O($^1$D)
states or non-Born-Oppenheimer effects.\cite{nikitin:book} Hence, an
increase of the computed rates $k^{\rm diss}(T)$ by one order of
magnitude only be adopting a more likely value for $g_{\rm e}$ is
conceivable.\\

\noindent
Finally, increasing the size of the basis set used, e.g. AVTZ to AVQZ,
will further improve the quality of the PESs. Within transition state
theory, a difference of $\sim 2$ orders of magnitude in the rate
points to differences in energies of 2 to 3 kcal/mol which is
reminiscent of the increase by 2.9 kcal/mol in the O$_2$ dissociation
energy when going from MRCI+Q/AVTZ to MRCI+Q/AVQZ, see above.\\

\subsection{Final State Distribution and NN-based model}
\begin{figure}[h!]
    \centering \includegraphics[width=\linewidth]{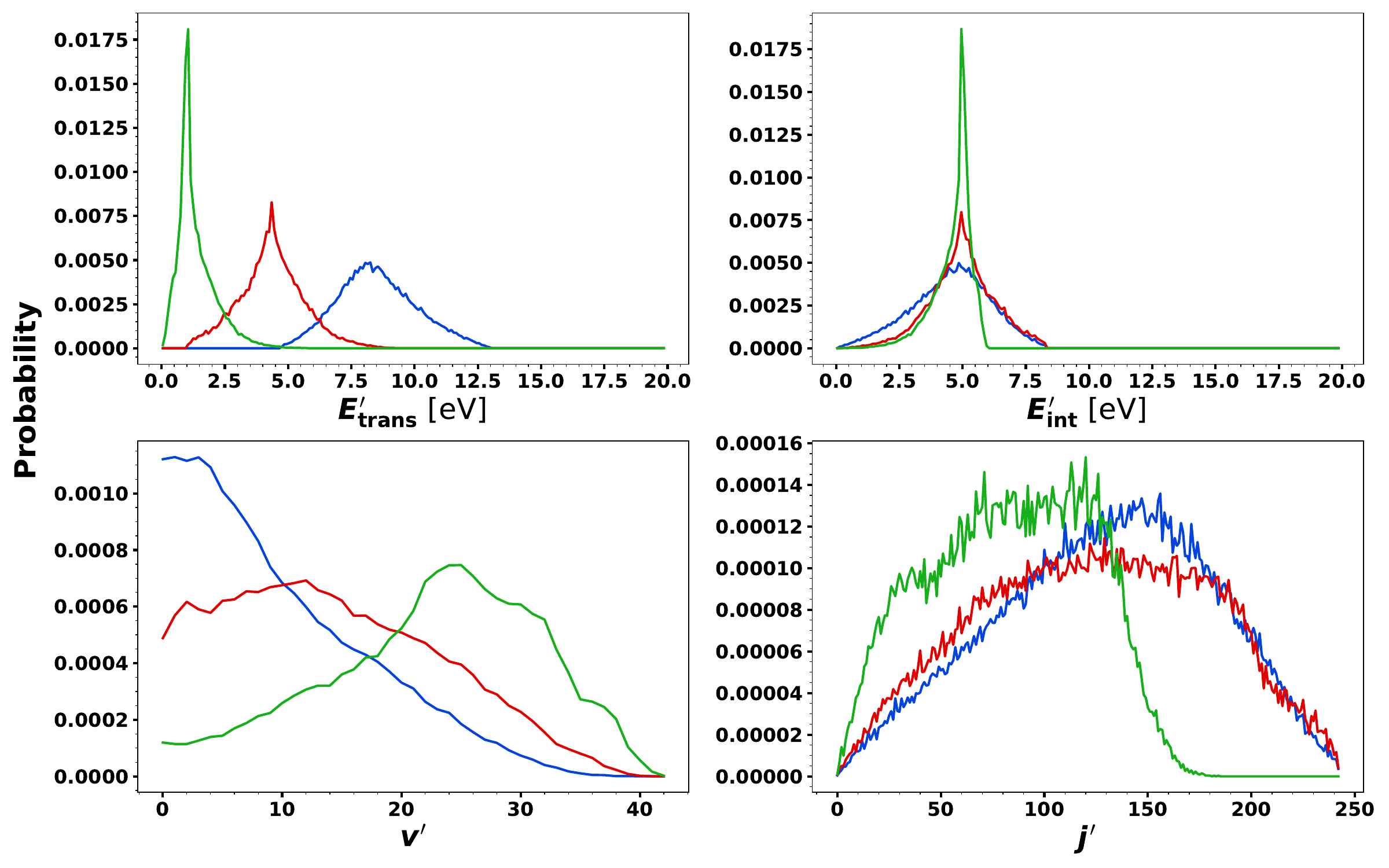}
    \caption{Product QCT distribution for the $^{1}$A$'$ RKHS-PES from
      3 different initial conditions: [$v=2,j=193, E_{\rm trans}=7.0$ eV]
      (blue), [$v=34,j=1, E_{\rm trans}=4.5$ eV] (red) and
      [$v=38,j=29, E_{\rm trans}=1.0$ eV] (green). The final $E_{\rm
        trans}'$, $E_{\rm int}'$, $v'$, and $j'$ are plotted as a
      function of the reaction probability. The probability is
      computed using histogram binning.}
    \label{fig:dist-detail}
\end{figure}

For hypersonic modeling, explicit computation of the state-to-state
rates from QCT simulations is rather time consuming. More
coarse-grained simulations benefit from machine-learned models based
on a small subset of the reactant states. In the following, a
state-to-distribution (STD) model and its performance for the atom
exchange reaction is described. First, representative final state
distributions $P(E_{\rm trans}')$, $P(E_{\rm int}')$, $P(v')$, and
$P(j')$ for the O($^3$P) + O$_2( ^3\Sigma_g^{-} )$ the reactions
O$_{\rm A}$ + O$_{\rm B}$O$_{\rm C}$ $\rightarrow$ O$_{\rm B}$ +
O$_{\rm A}$O$_{\rm C}$ or O$_{\rm C}$ + O$_{\rm A}$O$_{\rm B}$ are
considered, see Figure \ref{fig:dist-detail}. Depending on the initial
condition $[v,j,E_{\rm trans}]$ (see figure caption) the final state
distributions differ appreciably. This is due to the nonequilibrium
nature of the conditions under which the reaction occurs. For example,
with increasing initial $E_{\rm trans}$ the final $P(E_{\rm trans}')$
shifts to higher energies and broadens. On the other hand, for low
$E_{\rm trans}$ but high initial $v$, the final $P(j')$ decays to 0
for $j' < j_{\rm max}$ primarily due to full atomization as a
competition channel.\\

\begin{figure}[h!]
\centering \includegraphics[width=0.8\linewidth]{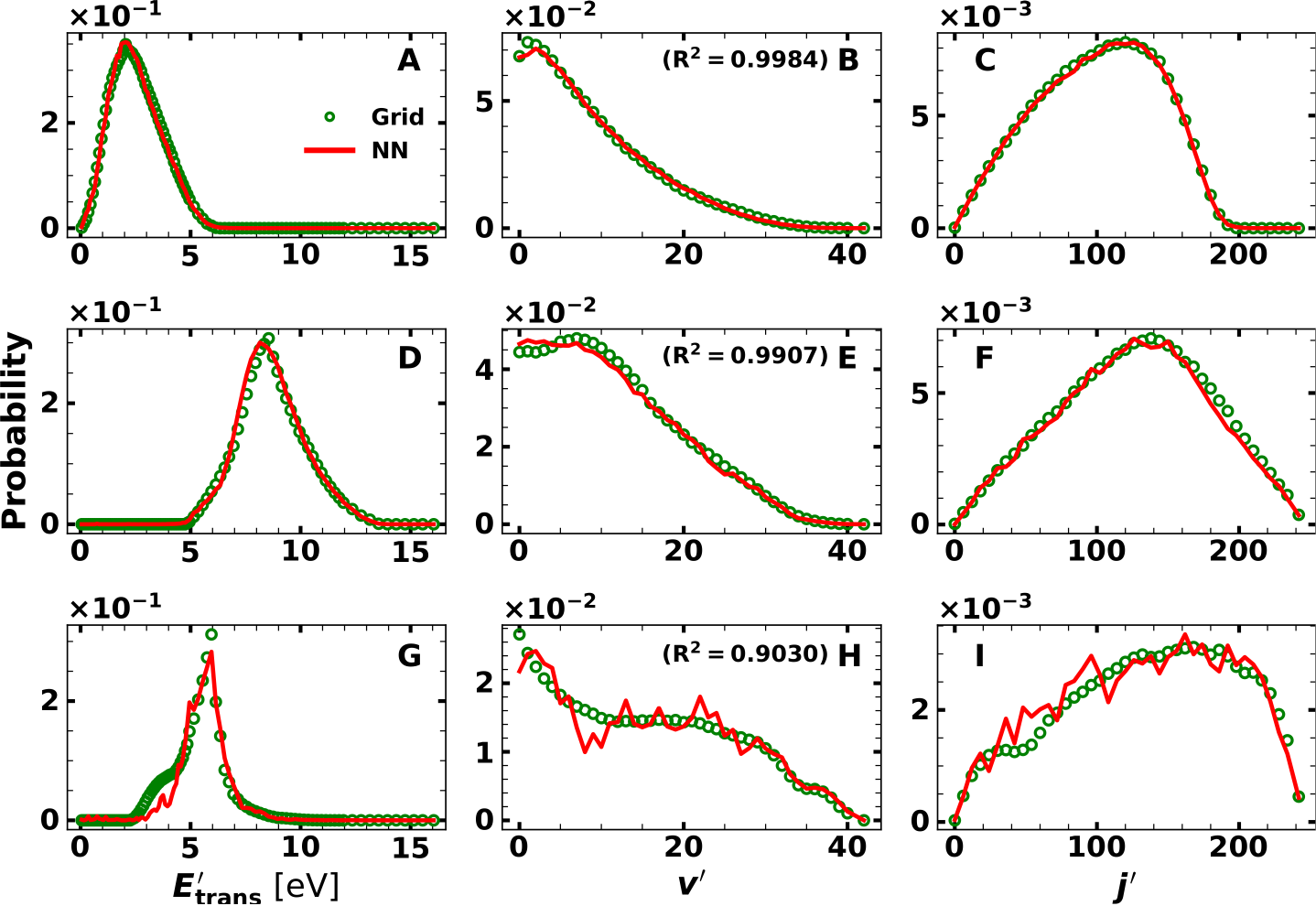}
\caption{NN-STD Model for final state distributions from QCT
  simulations using the RKHS-PES. Reference amplitudes (Grid) obtained
  from taking moving averages of the raw QCT data in comparison to STD
  predictions (NN). The final $E_{\rm trans}'$, $v'$, and $j'$ are
  plotted as a function of the reaction probability. Three cases for
  the quality of the NN-trained models are distinguished: initial
  condition for which the prediction is best (A to C, highest $R^2 =
  1.00$), is closest to the average $R^2$ (D to F, $R^2 = 0.99$), and
  worst (G to I, $R^2 = 0.69$). The corresponding initial conditions
  are [$E_{\rm trans} = 4.0$ eV, $v = 10$, $j = 29$]; [$E_{\rm trans}
    = 6.0$ eV, $v = 2$, $j = 221$]; [$E_{\rm trans} = 6.0$ eV, $v =
    38$, $j = 15$] for the best, mean and worst cases,
  respectively. Note that 34 samples were excluded from the dataset
  based on $\sum P_v<0.005$.}
\label{fig:nn-std}
\end{figure}

\begin{figure}
    \centering
    \includegraphics[width=\linewidth]{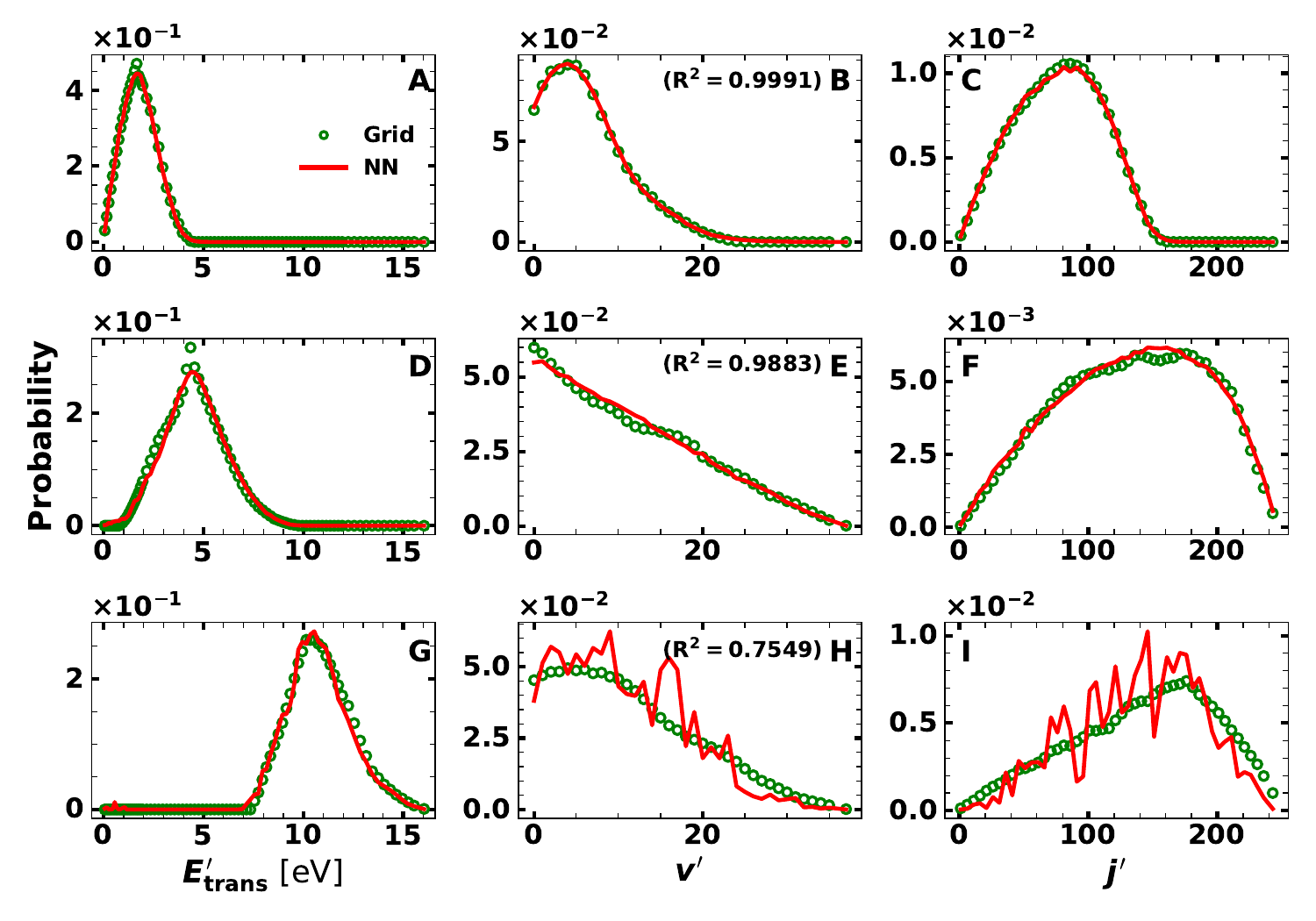}
    \caption{NN-STD Model for final state distributions from QCT
      simulations using the PIP-PES.  Reference amplitudes (Grid)
      obtained from taking moving averages of the raw QCT data in
      comparison to STD predictions (NN). The final $E_{\rm trans}'$,
      $v'$, and $j'$ are plotted as a function of the reaction
      probability. Three cases for the quality of the NN-trained
      models are distinguished: initial condition for which the
      prediction is best (A to C, highest $R^2 = 1.00$), is closest to
      the average $R^2$ (D to F, $R^2 = 0.99$), and worst (G to I,
      $R^2 = 0.77$). The corresponding initial conditions are [$E_{\rm
          trans} = 2.5$ eV, $v = 8$, $j = 29$]; [$E_{\rm trans} = 6.0$
        eV, $v = 18$, $j = 57$]; [$E_{\rm trans} = 8.0$ eV, $v = 2$,
        $j = 235$] for the best, average and worst cases,
      respectively.}
\label{fig:nn-std-truh}
\end{figure}

\noindent
Finally, STD models were trained following the procedures described in
the methods section for the RKHS and PIP-PESs, see Figures
\ref{fig:nn-std} and \ref{fig:nn-std-truh}. Both figures report the
performance of the trained model on off-grid initial conditions, which
were not used for training the NN. The top, middle, and bottom rows
report the best, average and worst predictions by the NN. Symbols
labelled ``grid'' refer to the reference amplitudes obtained from
running averages over the QCT simulations whereas the solid lines
represent the prediction from the NN. For the top two lines the
agreement between the NN-trained model and the true QCT simulations is
excellent whereas for the worst case (bottom row) $P(E_{\rm int}')$ is
still acceptable but for $P(v')$ and $P(j')$ the overall shape is
captured but details differ.\\

\noindent
Hence, the present work provides two statistical models for predicting
the entire state-to-state distributions based on rigorous QCT
simulations. Such NN-based models can be used in more coarse-grained
simulations of the reaction kinetics as it is, for example, done using
the PLATO (PLAsmas in Thermodynamic nOn-equilibrium)
software.\cite{munafo:2020}\\

\section{Discussion and Conclusions}
The present work presents, analyzes and uses a new RKHS-based reactive
PES for the O($^3$P) + O$_2( ^3\Sigma_g^{-} )$ collision system
suitable for applications to hypersonics. Validation of the PES is
done through extensive QCT simulations and comparison with thermal
rates for the exchange and atomization reactions at the respective
measurement conditions. In both cases the $T-$dependence of the rates
is correctly captured. Absolute rates are too low by about 2 orders of
magnitude for the atomization reaction. Reasons for this include the
level of theory at which the electronic structure calculations were
carried out and the degeneracy factor which can be expected to be
larger than 1/27 as used here and in previous work.\cite{boyd:2016}
The chosen level of theory (MRCI+Q/AVTZ) was primarily motivated by
the fact that with the present PES a full set of reactive and
validated PESs for the [NNO],\cite{MM.n2o:2020}
[NOO],\cite{MM.no2:2020} and [NNN],\cite{MM.n3:2024} reactive systems
is now available which can be used in more coarse-grained studies of
combustion processes and reaction networks. Further improvements, such
as the use of larger basis sets, are possible but are unlikely to
yield qualitatively different results.\\

\noindent
QCT simulations using an earlier PIP-PES are consistent with the
findings for $k(T)$ when running simulations with the RKHS
PES. Although the shapes of the PESs are related, the atom exchange
rates differ slightly which underlines the sensitivity of such
simulations to the topography and features of the PESs. Given that two
related PESs (RKHS and PIP) yield somewhat different thermal rates for
the exchange reaction motivates the question how to further improve
the PESs for yet better agreement with experiment. This can be
achieved for example by using morphing approaches guided by
experimental information.\cite{MM.morph:1999,MM.morph:2024} This has
been successfully done for the He--H$_2^+$ collision system by using
measurements of the H$_2^+$ translational spectrum.\\

\begin{figure}[h!]
    \centering
    \includegraphics[width=\linewidth]{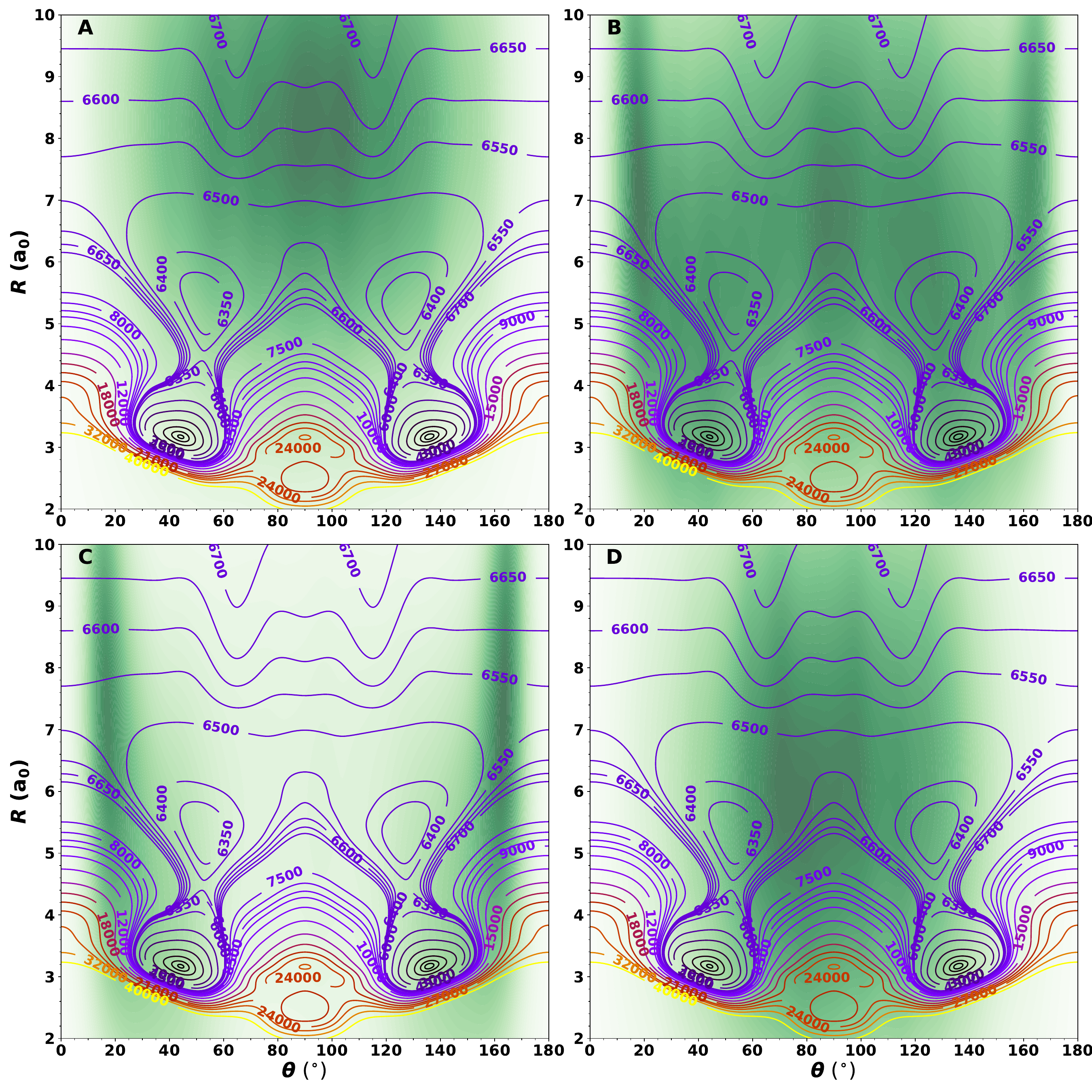}
    \caption{The cumulative probability distributions $P(R,\theta)$
      (black isocontours) from 1000 trajectories of each type
      (elastic, inelastic, atom exchange, atomization - panels A to D)
      projected onto the 2d-PES $V(R,\theta;r)$ for $r_{\rm OO} =
      2.282$ a$_0$. The distributions $P(R,\theta)$ are shown as
      normalized heatmaps with minimum and maximum amplitudes between
      [0.0, 1.0]. White to dark green regions refer to amplitudes of
      0.0 and 1.0, respectively.}
    \label{fig:prob}
\end{figure}

\noindent
To better understand how the process in question is related to
sampling the underlying PES, Figure \ref{fig:prob} reports the PES
$V(R,\theta)$ for a diatomic bond length $r_{\rm OO} = 2.282$ a$_0$
together with probability distributions $P(R,\theta)$ from 1000
QCT-trajectories for each of the four relevant processes: elastic (A),
inelastic (B), reactive (C), and dissociative (D) atom-plus-diatom
collisions. Elastic scattering (Figure \ref{fig:prob}A) probes
primarily the long-range part of the PES and does not sample the
region around the global minimum. Inelastic scattering (panel B), on
the other hand, samples the region around the global minimum and
migration along the two directions $\theta \sim 20^\circ$ and $\theta
\sim 160^\circ$ can be clearly identified. In addition, a
high-probability zone is also around $\theta \sim 90^\circ$. Contrary
to that, atom exchange reactions penetrate deeper into the region of
the two symmetry-related minima but do not sample the T-shaped
structure at all, see Figure \ref{fig:prob}C. Finally, atomization
trajectories approach the diatomic mainly along $\theta = 90^\circ$
and primarily probe the repulsive wall (Figure \ref{fig:prob}D), as
expected.\\

\noindent
Notably, both PESs used in the present work - RKHS and PIP - feature
``reefs'' in the entrance channel, see Figure
\ref{fig:pes-comp-m}. Nevertheless, the QCT-simulations using both
PESs reproduce the experimentally observed negative $T-$dependence of
the thermal rates for the exchange reaction. Hence, the notion that
the ``reef'' in earlier PESs is responsible for the positive
$T-$dependence of $k(T)$, which is inconsistent with experimental
observations, is not supported by the present work. For further
validation of the RKHS-PES quantum bound state calculations are
envisaged for which experimental data is also available for
comparison.\cite{chang:1994}\\

\noindent
With the present RKHS-PES all relevant atom + diatom reactions for
burning air are now described at a uniformly high level of quantum
chemical theory (MRCI+Q/AVTZ) with all PESs represented using
RKHS. The PESs for all systems [NOO], [NNO], [NNN], and [OOO] were
validated vis-a-vis thermal rates for the exchange and atomization
reactions $k^{\rm exch} (T)$ and $k^{\rm diss} (T)$,
respectively. This provides a unified framework to investigate the
reaction network using more coarse-grained approaches, for example.\\

\noindent
As evidenced here, further improvements can be envisaged, for example
through transfer learning\cite{MM.tl:2023} of the PESs to higher
levels of theory, such as MRCI+Q/AVQZ. On the other hand, the present
work and earlier investigations of the [NNN] reaction system provide
evidence that different levels of quantum chemical theory, e.g. CASPT2
instead of MRCI+Q, or larger basis sets do not lead to fundamental
changes in the rates, specifically for high-energy collisions. This
may be different, though, for reactions at low temperatures for which
additional care needs to be exercised.\cite{MM.n2o:2023} \\

\noindent
In summary, the thermal rates for the atom exchange and atomization
reactions using QCT simulations validate the RKHS representation of
the PES. Although both surfaces considered in the present work exhibit
more or less pronounced ``reef" structure in the long-range part of
the PES, QCT simulations using them correctly capture the
experimentally observed negative $T-$dependence for the exchange
reaction. This is observed for two different representations of the
PES based on two different quantum chemical levels of theory, which
further corroborates the validity of the RKHS- and PIP-PESs.

\section*{Data Availability}
The codes and data for the present study are available from
\url{https://github.com/MMunibas/NN-STD-O3} upon publication.

\section*{Acknowledgment}
The authors gratefully acknowledge financial support from the AFOSR
under award number FA8655-21-1-7048, the Swiss National Science
Foundation through grants $200020\_219779$ (MM), $200021\_215088$
(MM), the NCCR-MUST (MM), and the University of Basel (MM). The
authors thank Prof. R. Dawes for providing source code and help with
the DLLJG-PES and Prof. G. Schatz for scientific correspondence.\\

\clearpage
\bibliography{ref}

\clearpage

\renewcommand{\thetable}{S\arabic{table}}
\renewcommand{\thefigure}{S\arabic{figure}}
\renewcommand{\thesection}{S\arabic{section}}
\renewcommand{\d}{\text{d}}
\setcounter{figure}{0}  
\setcounter{section}{0}  
\setcounter{table}{0}

\newpage

\noindent
{\bf SUPPORTING INFORMATION: High-Energy Reaction Dynamics of O$_{3}$}

\begin{figure}
    \centering \includegraphics[width=\textwidth]{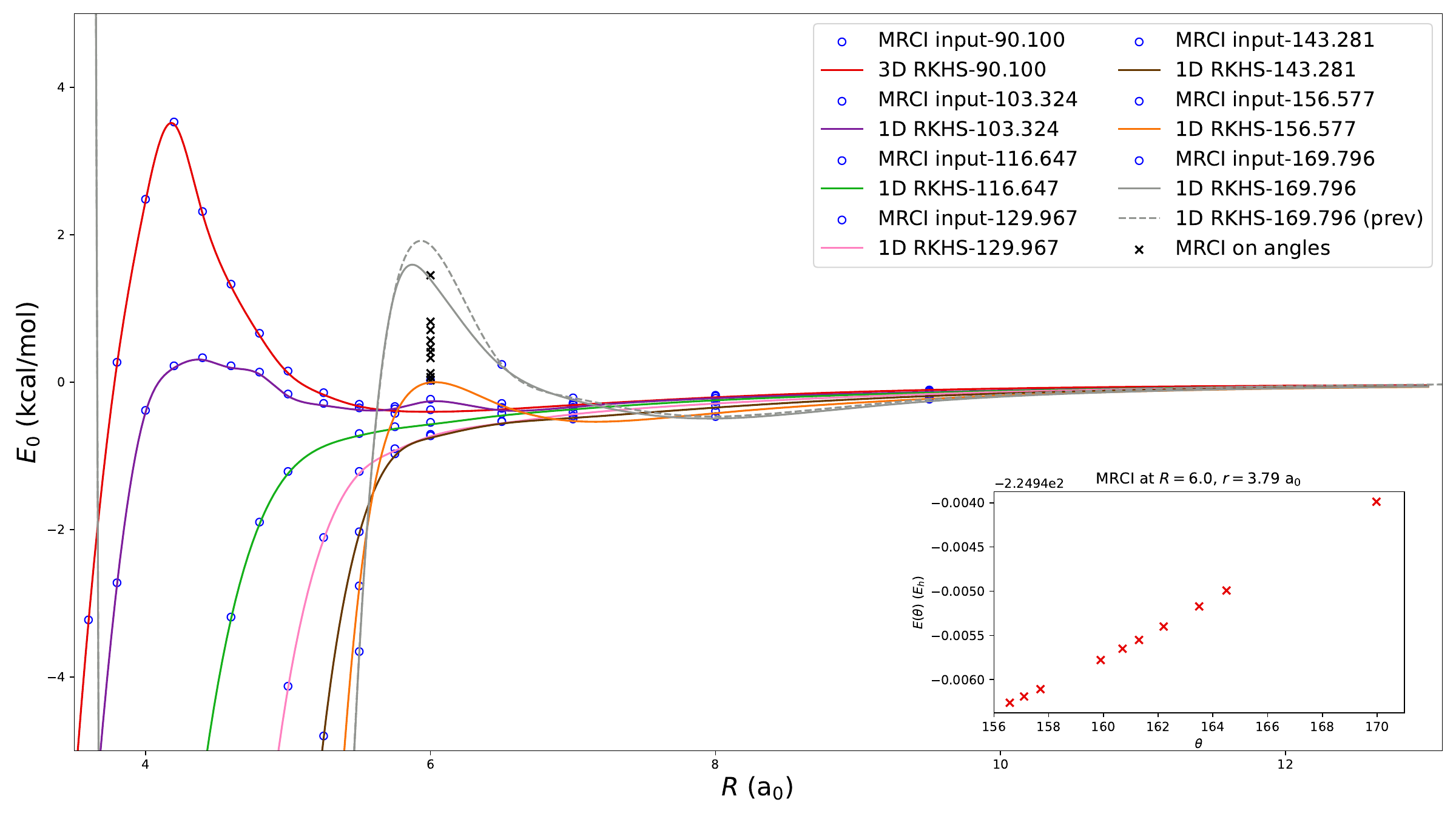}
    \caption{$V(R)$ of the 3D RKHS refitted with one extra training
      data compared to the MRCI training data at $169.796^{\circ}$ and
      $r=3.79$ a$_0$.}
    \label{sifig:comp-r3.79-refit}
\end{figure}

\begin{figure}
    \centering 
    \includegraphics[width=\textwidth]{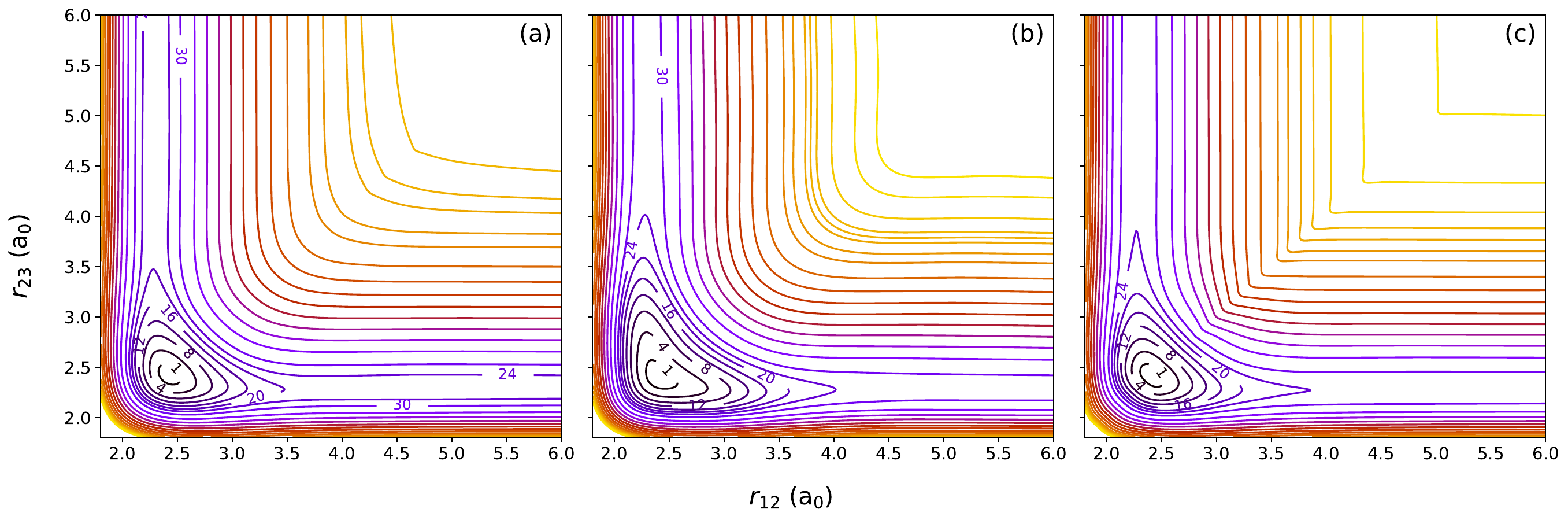}
    \caption{From left to right: $117^{\circ}$, $119^{\circ}$,
      $117^{\circ}$.} Contour of $V(r_{\rm AB},r_{\rm BC})$ between
    $[1.8,6.0]$ a$_0$ of all three PESs at their corresponding
    $\theta$ angles of tight minimum in valence coordinates. The
    subfigures from left to right: (a) RKHS-PES (b)
    PIP-PES;\cite{varga:2017} (c) PES3.\cite{Dawes:2011,Dawes:2013}
    \label{sifig:cont-compare}
\end{figure}

\begin{table}[ht]
    \centering
    \begin{tabular}{l|c|c|c}
\hline\hline
$T$(K) & $N_{\rm react}$ & $N_{\rm total}$ & Ratio \\
\hline
1000 & 0 & 5002227 & 0 \\
2000 & 0 & 5001402 & 0 \\
3000 & 0 & 4999216 & 0 \\
4000 & 6 & 5000171 & $1.20\times 10^{-6}$ \\
5000 & 134 & 5000606 & $2.68\times 10^{-5}$ \\
6000 & 816 & 4999019 & 0.00016 \\
8000 & 6627 & 5001339 & 0.0013 \\
10000 & 21310 & 5000949 & 0.0043 \\
15000 & 91174 & 5000299 & 0.018 \\
20000 & 177426 & 4998451 & 0.035 \\
\hline\hline
    \end{tabular}
    \caption{Statistics of the QCT results for dissociation reaction on RKHS PES}
    \label{sitab:ratio_disso_RKHS}
\end{table}

\begin{table}[ht]
    \centering
    \begin{tabular}{l|c|c|c}
\hline\hline
$T$(K) & $N_{\rm react}$ & $N_{\rm total}$ & Ratio \\
\hline
1000 & 0 & 5000800 & 0 \\
2000 & 0 & 5001996 & 0 \\
3000 & 0 & 5000755 & 0 \\
4000 & 2 & 5000152 & $4.00\times 10^{-7}$ \\
5000 & 85 & 5000915 & $1.70\times 10^{-5}$ \\
6000 & 514 & 5000549 & 0.00010 \\
8000 & 4529 & 5002607 & 0.00091 \\
10000 & 15821 & 4998037 & 0.0032 \\
15000 & 76869 & 4997764 & 0.015 \\
20000 & 156768 & 5000173 & 0.031 \\
\hline\hline
    \end{tabular}
    \caption{Statistics of the QCT results for dissociation reaction on PIP PES}
    \label{sitab:ratio_disso_PIP}
\end{table}

\end{document}